\newcommand{\qsla}{q\hspace{-0.45em}/}

\documentclass[english]{article}
\usepackage[T1]{fontenc}
\usepackage[latin9]{inputenc}
\usepackage{geometry}
\usepackage{amsmath}
\usepackage{graphicx}



\usepackage{babel}

\begin{document}

\global\long\def\dd{\mathrm{d}}


\title{Photon-photon scattering: a tutorial\vspace*{-2.5cm}\begin{flushright}
\normalsize Alberta Thy 12-11
\end{flushright}
\vspace*{1cm}
}

\author{Yi Liang and Andrzej Czarnecki}
\date{}

\maketitle

\begin{center}
\emph{Department of Physics, University of Alberta, Edmonton, Alberta,
Canada T6G 2E1}
\par\end{center}
\begin{abstract}
Long-established results for the low-energy photon-photon scattering,
$\gamma\gamma\to\gamma\gamma$, have recently been questioned.  We
analyze that claim and demonstrate that it is inconsistent with experience.
 We demonstrate that the mistake originates from an erroneous manipulation
of divergent integrals and discuss the connection with another recent
claim about the Higgs decay into two photons.  We show a simple way
of correctly computing the low-energy $\gamma\gamma$ scattering.
\end{abstract}


\section{Introduction}

After Dirac proposed the theory of negative energy solutions of his
equation \cite{Dirac:1930ek}, it was realized that photons can interact
with other photons by polarizing the vacuum. Photon-photon scattering
was qualitatively considered in this context by Halpern \cite{Halpern33},
and its cross section, for the case of photon energies low compared
to the electron mass, was determined by Euler and Kockel in 1935 \cite{EulerKockel35,Euler36}.
If the energy of each of the colliding photons is $\omega$ in the
frame in which their total momentum vanishes, the low-energy differential
cross section is \begin{equation}
\frac{d\sigma}{d\Omega}=\frac{139\alpha^{4}}{\left(180\pi\right)^{2}}\frac{\omega^{6}}{m^8}\left(3+\cos^{2}\theta\right)^{2},
\label{eq:correctCross}\end{equation}
where $\alpha\simeq1/137$ is the fine structure constant and $m$
is the electron mass. 

High energy scattering was considered soon afterward \cite{Akhi36,Akhi37}.
A thorough analysis of the scattering at all energies, including partial
cross sections for various polarization states, was carried out in
\cite{KarplusNeuman1,KarplusNeuman2}, using the then new diagrammatic
technique of Feynman. Since then, the photon-photon scattering cross
section has been confirmed with other methods, and even higher-order
QED corrections have been computed \cite{Dittrich:2000zu}. Results
obtained up to 1971 are reviewed in \cite{Costantini:1971cj} and
more recent developments are summarized in \cite{Martin:2003gb}.

Very recently, the classic result for the low-energy cross section
(\ref{eq:correctCross}) has been questioned \cite{Kanda:2011vu,Fujita:2011rd}.
In those papers, the cross section is found to be many orders of magnitude
larger, since it is not suppressed by powers of $\left(\omega/m\right)$,
but is proportional to $1/\omega^{2}$,\begin{equation}
\frac{d\sigma_{\mbox{{\scriptsize FK}}}}{d\Omega}=\frac{\alpha^{4}}{\left(12\pi\right)^{2}\omega^{2}}\left(3+2\cos^{2}\theta+\cos^{4}\theta\right).\label{eq:CrossFK}\end{equation}
As we will demonstrate in this paper, this claim is incorrect. It
has already been pointed out \cite{Bernard:2011vz} that it contradicts
existing laboratory bounds on the the photon-photon cross section,
obtained by colliding laser beams. We show in addition that a cross
section increasing with the inverse squared energy of the colliding
photons limits the mean free path of visible light due to collisions
with the cosmic microwave background radiation (CMBR) to less than
the distance between Earth and Jupiter. Thus the fact that we can
sharply see much more distant astronomical objects proves that the
low-energy photon-photon scattering must be significantly suppressed,
as predicted by eq. (\ref{eq:correctCross}). 

The matrix element for the photon-photon scattering is absent at the
tree level since photons are neutral. It arises only at the loop level.
The sum of all contributing loop diagrams must be finite since there
is no parameter in the QED Lagrangian whose renormalization could
absorb a divergence. Refs.~\cite{Kanda:2011vu,Fujita:2011rd} found
an incorrect result because of the assumption that if the sum of those
diagrams is finite, they can be calculated without any regularization.
In fact, even though the sum of the diagrams is finite, each of them
separately is divergent. Calculating the sum is somewhat delicate
and is easiest done with regularized loop integrals (see, however,
an alternative calculation in \cite{Schwinger:II} and another point
of view on avoiding regularization in \cite{Jackiw:1999qq}).

Interestingly, a similar error \cite{Gastmans:2011ks,Gastmans:2011wh}
has recently cast doubt over the rate of the Higgs boson decay into
two photons. That process, too, is loop induced, and the sum of contributing
loops is finite. But individual loop integrals are divergent and must
be regularized, as has already been thoroughly discussed in this context
\cite{Shifman:2011ri,Huang:2011yf,Marciano:2011gm,Jegerlehner:2011jm,Shao:2011wx}.

\section{Mean free path of photons in a microwave background}

The CMBR is a gas of photons with the spectrum of a black body at
a temperature of $1/\beta=2.725$ K. Here we want to compute how far
a visible-light photon with energy $E_{\gamma}\simeq2.5$ eV can travel
in such a gas before scattering, from the point of view of an observer
in whose frame the CMBR is isotropic. (We will call it the LAB frame.
For the purposes of this discussion an Earth-based observer is a good
approximation.) Consider one mode of the CMBR radiation, characterized
by its energy $E$ and inclination angle $\theta$ with respect to
the direction from which the visible photon is incident. The relative
velocity of the two photons (as seen in the LAB frame) is $\vec{v}_{1}+\vec{v}_{2}=\left(1+\cos\theta,-\sin\theta\right)$,
$\left|\vec{v}_{1}+\vec{v}_{2}\right|=2\cos\frac{\theta}{2}$ (we
use the units $c=\hbar=k_{B}=1$). In the frame where the total momentum
of the photons vanishes, each has the energy $\omega$ given by\begin{equation}
\omega=\frac{1}{2}\sqrt{2EE_{\gamma}\left(1+\cos\theta\right)}=\cos\frac{\theta}{2}\sqrt{EE_{\gamma}}.\label{eq:omega}\end{equation}
That energy determines the scattering cross section. Collisions with
photons in this particular mode will occur at the rate\begin{equation}
\mbox{d}\Gamma_{E\theta}=\left|\vec{v}_{1}+\vec{v}_{2}\right|\sigma\mbox{d}\rho\left(E\right)\label{eq:partialRate}\end{equation}
where \begin{equation}
\mbox{d}\rho\left(E\right)=\frac{E^{2}}{2\pi^{2}}\frac{\mbox{d}E\mbox{d}\cos\theta}{\exp\left(\beta E\right)-1}\label{eq:density}\end{equation}
is the density of CMBR photons with energy $E$, and $\sigma$ is
the scattering cross section. Integrating over the energies and directions
of the CMBR photons we find the mean free path. Between collisions,
the visible-light photon will travel on average the distance \begin{equation}
\lambda=\pi^{2}\left[\int_{0}^{\infty}\mbox{d}E\int_{-1}^{1}\mbox{d}\cos\theta\cos\frac{\theta}{2}\frac{E^{2}\sigma}{\exp\left(\beta E\right)-1}\right]^{-1}.\label{eq:meanFreePath}\end{equation}
We now consider the two formulas for the low-energy cross section.
If we use the classical result (\ref{eq:correctCross}), we find the
total cross section\begin{equation}
\sigma\left(\gamma\gamma\to\gamma\gamma\right)=\frac{973\alpha^{4}\omega^{6}}{10125\pi m^{8}},\label{eq:totalCrossCorrect}\end{equation}
and the mean free path\begin{eqnarray}
\lambda & = & \pi^{2}\left[\frac{973\alpha^{4}E_{\gamma}^{3}}{10125\pi m^{8}}\int_{-1}^{1}\mbox{d}\cos\theta\cos^{7}\frac{\theta}{2}\int_{0}^{\infty}\mbox{d}E\frac{E^{5}}{\exp\left(\beta E\right)-1}\right]^{-1}\nonumber \\
 & = & \pi^{2}\left[\frac{973\alpha^{4}E_{\gamma}^{3}}{10125\pi m^{8}}\cdot\frac{4}{9}\cdot\frac{8\pi^{6}}{63\beta^{6}}\right]^{-1}=\frac{820125m^{8}\beta^{6}}{4448\pi^{3}\alpha^{4}E_{\gamma}^{3}}.\label{eq:freePathClassic}\end{eqnarray}
Using $m=0.511$ MeV we find $\lambda\simeq7\cdot10^{68}$ meters,
a distance that would take light about $10^{43}$ times more time
to travel than the age of the Universe. In other words, the CMBR is
a rather transparent medium at visible frequencies. 

However, if we take instead the cross section suggested in \cite{Kanda:2011vu,Fujita:2011rd},
we find from eq.~\eqref{eq:CrossFK}\begin{equation}
\sigma_{\mbox{\scriptsize FK}}=\frac{29\alpha^{4}}{540\pi\omega^{2}},\label{eq:totalFK}\end{equation}
which gives a much shorter mean free path,

\begin{eqnarray}
\lambda_{\mbox{\scriptsize FK}} & = & \pi^{2}\left[\frac{29\alpha^{4}}{540\pi E_{\gamma}}\int_{-1}^{1}\frac{\mbox{d}\cos\theta}{\cos\frac{\theta}{2}}\int_{0}^{\infty}\mbox{d}E\frac{E}{\exp\left(\beta E\right)-1}\right]^{-1}\nonumber \\
 & = & \pi^{2}\left[\frac{29\alpha^{4}}{540\pi E_{\gamma}}\cdot4\cdot\frac{\pi^{2}}{6\beta^{2}}\right]^{-1}=\frac{810\pi\beta^{2}E_{\gamma}}{29\alpha^{4}},\label{eq:freePathFK}\end{eqnarray}
or $\lambda_{\mbox{\scriptsize FK}}=3\cdot10^{11}$ meters, equivalent
to about 15 light minutes. For comparison, the orbital radius of Jupiter
is about $8\cdot10^{11}$ meters. If the mean free path of the visible
light were so much shorter than even the radius of Jupiter's orbit,
no stars would be visible on the night sky. Clearly, the result eq.~\eqref{eq:CrossFK}
is at odds with experience.

The situation with eq.~\eqref{eq:CrossFK} is actually even worse.
Since the cross section is not suppressed by the mass of the electron,
there would be additional positive contributions from other charged
fermions that would differ only by the coupling constant, and would
further decrease the mean free path. This lack of suppression by the
inverse mass of the loop particle contradicts the Appelquist-Carazzone
decoupling theorem \cite{Appelquist:1974tg}.

\section{Determination of the photon-photon scattering}

\begin{figure}[h]
\centering\includegraphics[scale=0.3]{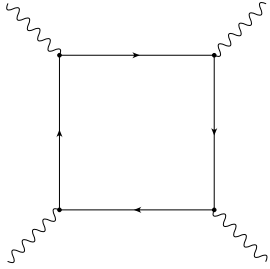}

\caption{Virtual electron loop inducing the four-photon coupling. \label{fig:Virtual-electron-loop}}

\end{figure}

In this section we present a derivation of the photon-photon scattering
matrix element in two regularization schemes: dimensional and Pauli-Villars.
We consider the box diagram shown in Fig.~\ref{fig:Virtual-electron-loop}.
External photons carry momenta $k_{1},\ldots,k_{4}$ which we will
consider as incoming, $k_{1}+k_{2}+k_{3}+k_{4}=0$.

There are six ways in which the four momenta can be arranged around
the oriented electron loop. However, diagrams that differ only by
the direction of the electron line give identical results so it is
enough to compute three of them, corresponding to three cyclic permutations
of $k_{1},k_{2},$ and $k_{3}$. (If there was an odd number of photons
coupling to the electron loop, the diagrams differing by the direction
of the electron would cancel one other, resulting in a vanishing amplitude.
This is the theorem due to Furry \cite{FurryTheorem}.)

At low energies of external photons it is especially easy to compute
the diagrams in Fig.~\ref{fig:Virtual-electron-loop}. We simply
Taylor-expand the electron propagators in the external momenta, so
that each propagator's denominator becomes simply 
$\left(\qsla-m\right)^{-1}=\left(\qsla+m\right)/\left(q^{2}-m^{2}\right)$
where $q$ is the loop momentum. Such expansion does not lead to any
spurious divergences, and commutes with the integration over $q$.
We now explain how this integration is performed in two regularization
schemes.

\subsection{Dimensional regularization}

Now that $q$ is present in the denominators only through $q^{2}$,
also in the numerator we can replace all scalar products of $q$ with
other vectors by powers of $q^{2}$ times products not involving $q$,\begin{equation}
q^{\mu_{1}}\ldots q^{\mu_{2n}}\to\frac{\Gamma\left(\frac{D}{2}\right)}{2^{n}\Gamma\left(\frac{D}{2}+n\right)}\left(q^{2}\right)^{n}S\left(g^{\mu_{1}\mu_{2}}\ldots g^{\mu_{2n-1}\mu_{2n}}\right).\label{eq:average}\end{equation}
Here $D$ is the space-time dimension and $S\left(g^{\mu_{1}\mu_{2}}\ldots g^{\mu_{2n-1}\mu_{2n}}\right)$
is a sum of products of $n$ metric tensors $g$, totally symmetric
in all indices $\mu_{i}$; it has $\left(2n-1\right)!!$ terms. Terms
odd in $q$ vanish upon integration.

The powers of $q^{2}$ resulting from \eqref{eq:average} can be canceled
against the denominators and the loop integration can be completed
using\begin{equation}
\int\frac{\dd^{D}q}{\left(2\pi\right)^{D}}\frac{1}{\left(q^{2}-m^{2}+i0\right)^{a}}=\frac{\left(-1\right)^{a}i}{\left(4\pi\right)^{D/2}}m^{D-2a}\frac{\Gamma\left(a-\frac{D}{2}\right)}{\Gamma\left(a\right)}.\label{eq:loopInt}\end{equation}
Each of the three diagrams contains terms with the exponent $a=2$,
leading to a divergence $\Gamma\left(2-\frac{D}{2}\right)\sim1/\left(D-4\right)$.
The divergences cancel when we add all three contributions. But individual
diagrams containing singularities $1/\left(D-4\right)$ have also
$D$-dependent factors, arising from the averaging in eq. \eqref{eq:average}.
The resulting finite contributions \emph{do not} cancel among themselves. 

How do these remaining terms depend on $m$? We remember that they
arise from the $a=2$ sector, therefore they scale like $m^{0}$ (the
overall dimension of the $\gamma\gamma\to\gamma\gamma$ amplitude).
There are other terms that scale with this power, arising from convergent
integrals like $m^{2}\int\dd^{4}q/\left(q^{2}-m^{2}\right)^{3}$.
The essential point is that the sum of all $m^{0}$ terms, including
the remnants of singularities, adds up to zero. The total result for
the amplitude turns out to be suppressed by four powers of $1/m$.

\subsection{Pauli-Villars regularization}

Another way of carrying out this calculation is to stay in four dimensions
but add another amplitude, with the electron replaced by a very heavy
particle of mass $M$, and with an opposite sign than the electron
loop. The calculation proceeds very similarly to the case of dimensional
regularization, with two changes. In averaging over the loop momentum
directions \eqref{eq:average} we replace $\frac{\Gamma\left(D+2n-1\right)}{\Gamma\left(D\right)}$
by its value at $D=4$, $\left(2n+2\right)!/6$. The formula for the
loop integration \eqref{eq:loopInt} is also replaced by its $D=4$
value, except in the divergent case $a=2$. In the dimensional regularization,
this divergent integral gives $m^{D-4}\Gamma\left(2-\frac{D}{2}\right)\to\frac{2}{4-D}-\ln m^{2}$.
In the Pauli-Villars approach one finds a convergent combination\begin{equation}
\int\frac{\dd^{4}q}{\left(2\pi\right)^{4}}\left[\frac{1}{\left(q^{2}-m^{2}+i0\right)^{2}}-\frac{1}{\left(q^{2}-M^{2}+i0\right)^{2}}\right]=\frac{i}{16\pi^{2}}\ln\frac{M^{2}}{m^{2}}.\label{eq:log}\end{equation}
When the three diagrams are added, this logarithm cancels, but now
there are no finite remnants of the singularities. Instead, the $m^{0}$
terms from the convergent diagrams are canceled by the $M^{0}$ terms
from the Pauli-Villars subtraction. Since they are independent of
the electron mass, they are the same in the amplitudes with the electron
and with the very heavy particle, and cancel in the difference.

In both regularization schemes, the only remaining result is suppressed
by the electron mass.

\subsection{Potential error from neglecting regularization}

We have just seen that the regularization is crucial in computing
the photon-photon scattering amplitude, even though the final result
does not contain divergences. We now want to inspect more closely
the part of the amplitude that does not contain external photon momenta,
and thus scales like the zeroth power of the electron mass,

\begin{eqnarray}
{\cal M}_{m^{0}} & \sim & \int\frac{\dd^{D}q}{\left(q^{2}-m^{2}+i0\right)^{4}}\Big[m^{4}S_{1}^{\mu\nu\rho\sigma}+2m^{2}\left(2S_{2}^{\mu\nu\rho\sigma}-q^{2}S_{1}^{\mu\nu\rho\sigma}\right)\nonumber \\
 &  & +24q^{\mu}q^{\nu}q^{\rho}q^{\sigma}+\left(q^{2}\right)^{2}S_{1}^{\mu\nu\rho\sigma}-4q^{2}S_{2}^{\mu\nu\rho\sigma}\Big]\epsilon_{1\mu}\epsilon_{2\nu}\epsilon_{3\rho}\epsilon_{4\sigma},\label{eq:noPowerOfM}\end{eqnarray}
with

\begin{eqnarray*}
S_{1}^{\mu\nu\rho\sigma} & = & g^{\mu\nu}g^{\rho\sigma}+g^{\mu\rho}g^{\nu\sigma}+g^{\mu\sigma}g^{\rho\nu},\\
S_{2}^{\mu\nu\rho\sigma} & = & g^{\mu\nu}q^{\rho}q^{\sigma}+\mbox{five other terms, }\end{eqnarray*}

where the terms not shown in $S_{2}$ have the other five distributions
of indices so that both $S_{1}$ and $S_{2}$ are totally symmetric
in $\mu,\nu,\rho,\sigma$. The second line in \eqref{eq:noPowerOfM}
contains four powers of the loop momentum $q$ and thus represents
divergent integrals. Without regularization, these divergent integrals%
simply do not have a meaning. If we apply the averaging procedure
\eqref{eq:average} to these terms, we find \begin{eqnarray}
\left\langle S_{2}^{\mu\nu\rho\sigma}\right\rangle  & = & \frac{2q^{2}}{D}S_{1}^{\mu\nu\rho\sigma},\nonumber \\
\left\langle q^{\mu}q^{\nu}q^{\rho}q^{\sigma}\right\rangle  & = & \frac{\left(q^{2}\right)^{2}}{D\left(D+2\right)}S_{1}^{\mu\nu\rho\sigma},\label{eq:averagesS}\end{eqnarray}
so that if $D=4$, the second line of \eqref{eq:noPowerOfM} vanishes,
as does the term $\sim m^{2}$ in its first line. Thus, if the regularization
is neglected, one is left with only the first term $m^{4}S_{1}$ which,
after the $q$ integration, gives a result independent of the electron
mass, scaling like $m^{0}$,\begin{eqnarray}
i\mathcal{M}_{m^{0}} & = & -\frac{4}{3}\alpha^{2}S_{1}^{\mu\nu\rho\sigma}\epsilon_{1\mu}\epsilon_{2\nu}\epsilon_{3\rho}\epsilon_{4\sigma}\label{eq:amplWrongA}\\
 & = & -\frac{4}{3}\alpha^{2}\left(\epsilon_{1}\cdot\epsilon_{2}\,\epsilon_{3}\cdot\epsilon_{4}+\epsilon_{1}\cdot\epsilon_{3}\,\epsilon_{2}\cdot\epsilon_{4}+\epsilon_{1}\cdot\epsilon_{4}\,\epsilon_{2}\cdot\epsilon_{3}\right),\label{eq:amplWrong}\end{eqnarray}
where $\epsilon_{i}$ are the polarization vectors of the four photons.
This dependence of the amplitude only on the polarization vectors
(and not on the photon momenta) means that the induced coupling of
the photons involves only their vector potential (the induced
effective operator is proportional to $(A^2)^2$), and not its derivatives.
It is not possible to construct such a coupling in a gauge invariant
way. 

This violation of gauge invariance may also generate photon's mass.
For example, if two of the external photon lines in Fig.~\ref{fig:Virtual-electron-loop}
are contracted, the resulting two-loop diagram generates an operator
$\sim A^2$, thus giving the photon a mass.  

In order to see how the cross section in
eq.~\eqref{eq:CrossFK} 
follows from the amplitude \eqref{eq:amplWrong},
we define two transverse polarization vectors for each photon, $\vec{\epsilon}_{i}^{1,2}$,
with $\vec{\epsilon}^{1}$ perpendicular to the scattering plane and
$\vec{\epsilon}^{2}$ lying in that plane.   We do not include here the
longitudinal photon polarizations, present if the
photon becomes massive, even though they may dominate the
cross section; however, our goal here is merely to explain how the
result \eqref{eq:CrossFK}  is related to the gauge-invariance
violating amplitude \eqref{eq:amplWrong}.
For the eq. \eqref{eq:amplWrong}
to give a non-zero result, each polarization must be represented an
even number of times (otherwise there will always be a factor 0 in
every term). There are eight possible such polarization assignments,
giving the following values of the three terms in \eqref{eq:amplWrong},\begin{eqnarray*}
{\cal M}_{1111} & \sim & 1+1+1,\\
{\cal M}_{2222} & \sim & 1+\cos^{2}\theta+\cos^{2}\theta,\\
{\cal M}_{1122} & \sim & 1+0+0,\\
{\cal M}_{1212} & \sim & 0+\cos\theta+0,\\
{\cal M}_{1221} & \sim & 0+0+\cos\theta,\end{eqnarray*}
and the last three amplitudes enter with a weight factor of 2, due
to the symmetry $1\leftrightarrow2$. The various amplitudes differ
by the polarization of some photons, so they do not interfere. The
sum of their squares gives $3^{2}+\left(1+2\cos^{2}\theta\right)^{2}+2+4\cos^{2}\theta=4\left(3+2\cos^{2}\theta+\cos^{4}\theta\right)$,
the angular structure of the (incorrect) result quoted in \eqref{eq:CrossFK}.
In fact, a sum over the polarizations of the final state photons and
an average over the polarizations of the initial state photons, leads
to the cross section given in eq. \eqref{eq:CrossFK}.%

What went wrong in the above procedure? The formulas \eqref{eq:averagesS}
cannot be applied to the divergent integrals in the second line of
\eqref{eq:noPowerOfM} in $D=4$, without regularization. If we stay
in $D$ dimensions, the terms we found to be zero in the $D\to4$
limit given finite contributions that cancel against the first term
of the integrand, $\sim m^{4}S_{1}$. In this correct treatment the
resulting amplitude is suppressed by four powers of $1/m$. 

\begin{figure}[h]
\centering\includegraphics[scale=0.4]{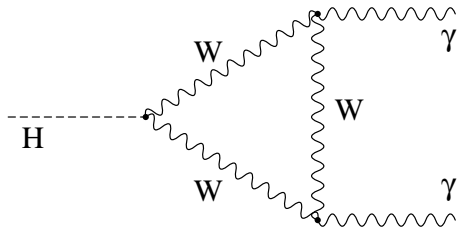}

\caption{An example of a $W$ boson loop mediating the Higgs boson decay into
two photons. \label{fig:Virtual-electron-loop-1}}
\end{figure}

The recent incorrect claim about the decay $H\to\gamma\gamma$
\cite{Gastmans:2011ks,Gastmans:2011wh} 
originated with a similar, but somewhat simpler integral. An example
of a contribution to that process is shown in Fig.~\eqref{fig:Virtual-electron-loop-1}.
There are only three propagators, and the divergent integrals are
present in the combination \cite{Gastmans:2011wh}\[
I_{\mu\nu}\left(D\right)=\int\dd^{D}q\frac{q^{2}g_{\mu\nu}-4q_{\mu}q_{\nu}}{\left(q^{2}-m^{2}+i0\right)^{3}}.\]
Without dimensional regularization, if we take $D=4$, it seems that
this integral vanishes after averaging over $q$ with help of \eqref{eq:average}.
As we have seen with the example of $\gamma\gamma$ scattering, and
as has already been discussed in the literature
\cite{Marciano:2011gm,Shifman:2011ri,Jackiw:1999qq},
such manipulations with unregulated, divergent integrals are unjustified.
In the case of the Higgs decay, they lead to the incorrect conclusion
that $I_{\mu\nu}\left(D\to4\right)$ vanishes. In fact, in the limit
of a very heavy Higgs boson, the correct finite result of $I_{\mu\nu}$
gives the most important contribution. 

\subsection{Results for polarized photons}

The correct result of the loop integration in the $\gamma\gamma\to\gamma\gamma$
amplitude contains scalar products among photon momenta $k_{i}$,
in addition to the polarization vectors $\epsilon_{i}^{\lambda}$.
The effective photon-photon coupling induced in this way is described
by operators involving the electromagnetic field tensor and is gauge
invariant. We know calculate the scattering cross sections for various
polarization situations. Instead of the linear polarizations we have
just considered, the scattering amplitudes will be presented in terms
of circular polarization states. Thus we introduce\[
\vec{\epsilon}^{\pm}=\frac{1}{\sqrt{2}}\left(\vec{\epsilon}^{1}\pm i\vec{\epsilon}^{2}\right),\]
describing right- and left-handed polarization states, respectively.
There are four possible initial polarization states, but it is sufficient
to consider just two of them, $++$ and $+-$. We get three independent
scattering amplitudes, ${\cal M}_{++++}$, ${\cal M}_{+++-}$, and
${\cal M}_{++--}$. We describe kinematics in terms of Mandelstam
variables $s=\left(k_{1}+k_{2}\right)^{2}=4E^{2}$ and $t=\left(k_{1}+k_{3}\right)^{2}=-2E^{2}\left(1-\cos\theta\right)$
and find

\begin{eqnarray}
i{\cal M}_{++++} & = & \frac{2\alpha^{2}\left(s^{2}+st+t^{2}\right)}{15m^{4}}=\frac{8\alpha^{2}\omega^{4}\left(3+\cos^{2}\theta\right)}{15m^{4}},\nonumber \\
i{\cal M}_{+++-} & = & -\frac{\alpha^{2}st\left(s+t\right)}{315m^{6}}=-\frac{16\alpha^{2}\sin^{2}\theta\omega^{6}}{315m^{6}},\nonumber \\
i{\cal M}_{++--} & = & -\frac{11\alpha^{2}s^{2}}{45m^{4}}=-\frac{176\alpha^{2}\omega^{4}}{45m^{4}}.\label{eq:amplits}\end{eqnarray}

We note that the amplitude ${\cal M}_{+++-}$ vanishes at the leading
order in $E/m$ expansion at which the other amplitudes are finite.
In order to compute it, we have to evaluate two more terms in the
Taylor expansion.

All other amplitudes can be obtained from eq. \eqref{eq:amplits}
using space and/or time reversal and the crossing symmetry. For example,
$i{\cal M}_{+-+-}=-\frac{11\alpha^{2}t^{2}}{45m^{4}}$ and $i{\cal M}_{+--+}=-\frac{11\alpha^{2}\left(s+t\right)^{2}}{45m^{4}}$

\subsection{Total cross section}

Once the polarized amplitudes have been evaluated, the unpolarized
cross section can be easily found. We quote here only the leading
low-energy result (thus we neglect ${\cal M}_{+++-}$ and seven amplitudes
related to it) for the cross section averaged over initial and summed
over final polarizations,\begin{eqnarray}
\frac{\dd\sigma\left(\gamma\gamma\to\gamma\gamma\right)}{\dd\Omega} & = & \frac{1}{256\pi^{2}\omega^{2}}\cdot\frac{\left|{\cal M}_{++++}\right|^{2}+\left|{\cal M}_{++--}\right|^{2}+\left|{\cal M}_{+-+-}\right|^{2}+\left|{\cal M}_{+--+}\right|^{2}}{2}\\
 & = & \frac{139\alpha^{4}\omega^{6}}{\left(180\pi\right)^{2}m^8}\left(3+\cos^{2}\theta\right)^{2},\end{eqnarray}
in agreement with the classic result \eqref{eq:correctCross}. When
integrated over both $\theta$ and $\phi$ from 0 to $\pi$ (we integrate
only over one hemisphere since the two final-state photons are identical),
this gives the total photon-photon scattering cross section,
\begin{equation}
\sigma\left(\gamma\gamma\to\gamma\gamma\right)=\frac{973\alpha^{4}\omega^{6}}{10125\pi
  m^8},
\end{equation}
in agreement with \cite{Landau4}. Other texts seem to have misprints
in these results \cite{IZ,ab65}.

\section{Conclusions}

We have demonstrated that the recently claimed result for the $\gamma\gamma\to\gamma\gamma$
scattering cross-section \eqref{eq:CrossFK} must be wrong. The photon-photon
coupling is induced by virtual loops with charged particles and is
suppressed at low photon energy by the inverse power of the electron
mass. The result \eqref{eq:CrossFK} lacks this suppression and yields
a very large cross-section, therefore a short mean free path of visible
photons even in the rare cosmic microwave radiation background. Such
short path would obscure all astronomical objects as close as Jupiter.

We have showed that the error resulted from manipulating unregulated
divergent integrals. A similar error misled the authors of \cite{Gastmans:2011ks,Gastmans:2011wh}
in the context of the Higgs decay to two photons. Both processes arise
only at the loop level and their amplitudes must be finite, since
there are no parameters in the Lagrangian that could absorb a divergence.
However, both processes are usually computed from a sum of several
diagrams, among which some are divergent. For this reason, a regularization
of individual contributions is necessary. 

Interestingly, the mistakes in these recent studies of
$\gamma\gamma\to\gamma\gamma$ and $H\to\gamma\gamma$ led to confusions
about various types of decoupling. The correct result for the former
process does respect Appelquist-Carazzone decoupling theorem in the
limit of low photon energies or large electron mass, whereas the
incorrect result of \cite{Kanda:2011vu,Fujita:2011rd} does not. On the
other hand, the correct result for the Higgs decay does not vanish, as
one could naively expect, in the limit of large Higgs mass
\cite{Shifman:2011ri} (or, equivalently, low $W$ boson mass; this type
of decoupling affects for example quarks but not the longitudinal $W$
components), while part of the reason why
\cite{Gastmans:2011ks,Gastmans:2011wh} believed their result was that
it did vanish in that limit.

We have also showed how the $\gamma\gamma\to\gamma\gamma$ amplitude
can be calculated in the low-energy regime, and how an expansion in
powers of the photon energy to the electron mass ratio can be organized.
This tutorial illustrates useful techniques of loop calculations:
averaging over loop momentum direction, loop momentum integration,
and various regularizations. We hope it will be helpful for other
similar loop calculations.

\subsection*{Acknowledgment}

We gratefully acknowledge helpful discussions with Alexander Penin and
Arkady Vainshtein,
and support of this research by Science and Engineering Research Canada
(NSERC). We thank Ben O'Leary for pointing out typographical mistakes
in the first version of this paper. 


\end{document}